\documentclass[sigconf]{acmart}

\AtBeginDocument{%
  \providecommand\BibTeX{{%
    \normalfont B\kern-0.5em{\scshape i\kern-0.25em b}\kern-0.8em\TeX}}}

\copyrightyear{2022}
\acmYear{2022}
\setcopyright{acmlicensed}\acmConference[CHI '22]{CHI Conference on Human Factors in Computing Systems}{April 29-May 5, 2022}{New Orleans, LA, USA}
\acmBooktitle{CHI Conference on Human Factors in Computing Systems (CHI '22), April 29-May 5, 2022, New Orleans, LA, USA}
\acmPrice{15.00}
\acmDOI{10.1145/3491102.3501829}
\acmISBN{978-1-4503-9157-3/22/04}

\usepackage{listings}
\usepackage[nameinlink]{cleveref}
\usepackage{subfig}
\usepackage{color, soul}
\usepackage[font={scriptsize,sf},tableposition=top]{caption}
\usepackage{float}
\floatstyle{plaintop}
\restylefloat{table}
\usepackage{sparklines}
\usepackage{makecell}
\usepackage{multirow}
\usepackage{multicol}
\usepackage{colortbl}
\usepackage{pifont}
\usepackage{tabularx}
\usepackage{balance}
\usepackage{amsmath}
\usepackage[utf8]{inputenc}
\usepackage[T1]{fontenc}

\usepackage{array}

\begin{document}

\title{Shape-Haptics: Planar \& Passive Force Feedback Mechanisms for Physical Interfaces}

\author{Clement Zheng}
\email{clement.zheng@nus.edu.sg}
\affiliation{
  \department{Division of Industrial Design}    
  \institution{National University of Singapore}
  \city{Singapore}
  \country{Singapore}
}

\author{Zhen Zhou Yong}
\email{e0325771@u.nus.edu}
\affiliation{
  \department{Division of Industrial Design}    
  \institution{National University of Singapore}
  \city{Singapore}
  \country{Singapore}
}

\author{Hongnan Lin}
\email{hlin324@gatech.edu}
\affiliation{
  \department{Industrial Design}    
  \institution{Georgia Institute of Technology}
  \city{Atlanta, Georgia}
  \country{USA}
}

\author{HyunJoo Oh}
\email{hyunjoo.oh@gatech.edu}
\affiliation{
  \department{Industrial Design \& Interactive Computing}    
  \institution{Georgia Institute of Technology}
  \city{Atlanta, Georgia}
  \country{USA}
}

\author{Ching Chiuan Yen}
\email{didyc@nus.edu.sg}
\affiliation{
  \department{Division of Industrial Design \& Keio-NUS CUTE Center}    
  \institution{National University of Singapore}
  \city{Singapore}
  \country{Singapore}
}

\renewcommand{\shortauthors}{Zheng, et al.}

\newcommand{\name}[0]{\textit{Shape-Haptics}}

\newcolumntype{L}[1]{>{\raggedright\let\newline\\\arraybackslash\hspace{0pt}}m{#1}}
\newcolumntype{C}[1]{>{\centering\let\newline\\\arraybackslash\hspace{0pt}}m{#1}}

\begin{teaserfigure}
    \centering
    \includegraphics[width=\textwidth]{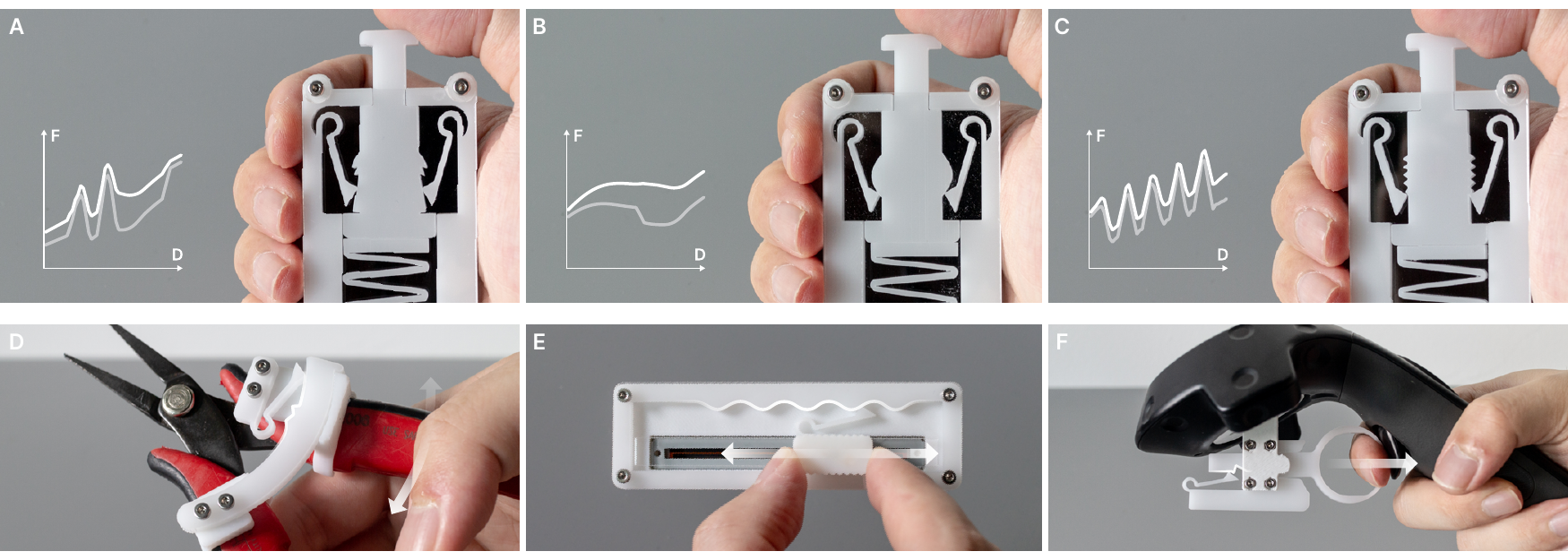}
    \vspace*{-7mm}
    \Description[Example Shape-Haptics objects]{Top, three Shape-Haptics linear mechanism swatches with different resulting force displacement curves. Below, three example applications of Shape-Haptics mechanisms: A pair of pliers that ``click'' when squeezed, electronic linear sliding potentiometer with a wave like profile that delivers variable resistance when sliding, and a VR controller trigger with an additional mid-point click.}
    \caption{A, B, C: \name\ mechanism swatches and their respective force-displacement curves. Objects with haptic attachments---D: Pliers that deliver a ``click'' at the end of a press. E: Electronic slider that delivers variable resistance along its track. F: Analog trigger on a VR controller augmented with a midpoint bump.}
    \label{fig:intro}
\end{teaserfigure}


\begin{abstract}
We present \name, an approach for designers to rapidly design and fabricate passive force feedback mechanisms for physical interfaces. Such mechanisms are used in everyday interfaces and tools, and they are challenging to design. \name\ abstracts and broadens the haptic expression of this class of force feedback systems through 2D laser cut configurations that are simple to fabricate. They leverage the properties of polyoxymethylene plastic and comprise a compliant spring structure that engages with a sliding profile during tangible interaction. By shaping the sliding profile, designers can easily customize the haptic force feedback delivered by the mechanism. We provide a computational design sandbox to facilitate designers to explore and fabricate \name\ mechanisms. We also propose a series of applications that demonstrate the utility of \name\ in creating and customizing haptics for different physical interfaces.
\end{abstract}



\maketitle
\section{Introduction}
\label{sec:introduction}

Haptics plays an important role in our interactions with physical interfaces and tools everyday. It is consequently a key consideration when designing physical interfaces. However, as identified by many HCI researchers \cite{van_oosterhout_facilitating_2020, schneider_haptic_2017, oviatt_multisensory_2017, seifi_haptipedia_2019}, it is also challenging to design for haptics. As researchers working in the intersection of tangible interactions and industrial design, we are motivated to facilitate designing haptics for physical interfaces---and broaden their range of haptic expression.

\begin{figure}[h]
    \centering
    \includegraphics[width=\linewidth]{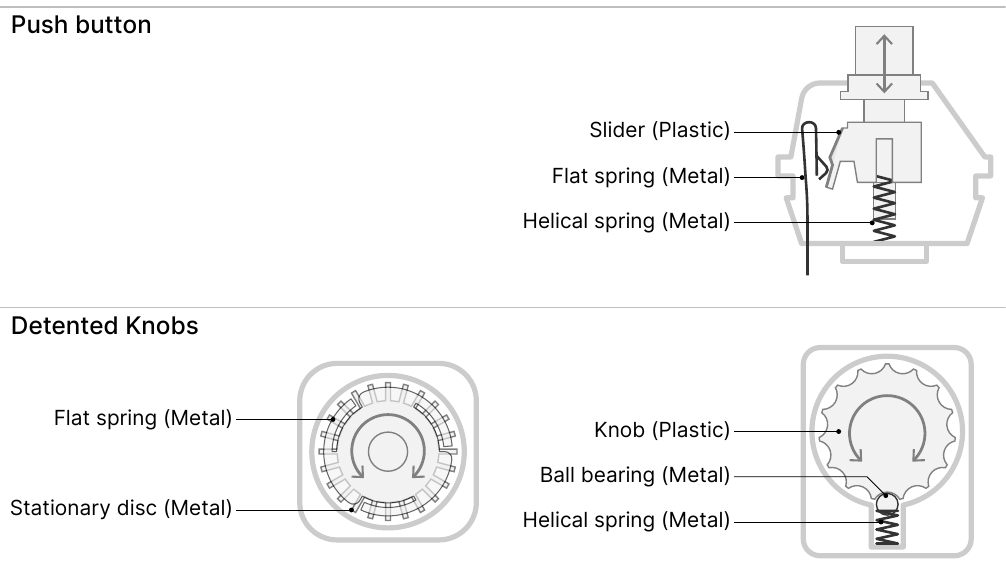}
    \Description[Illustrations of existing passive force feedback mechanisms]{Illustrations of existing passive force feedback mechanisms. A keyboard push button is a plastic slider that clicks when it moves against a flat spring and helical spring. Two types of detented knobs: type A comprises a flat spring that brushes against the bumps on a stationary metal disk, type B comprises a spring-loaded metal ball bearing that movies against a plastic knob with grooves.}
    \caption{Illustrations of existing passive force-feedback mechanisms.}
    \label{fig:existing}
    \vspace*{-3mm}
\end{figure}

The haptics delivered by physical interfaces and tools are typically generated by passive (unpowered) mechanisms. These mechanisms deliver force feedback in response to tangible interaction. Despite the ubiquity of such mechanisms, they are challenging to design as they often comprise complicated 3D assemblies that couple force feedback with other functions. For example, a tactile push button's mechanism supports opening and closing of electrical contacts and delivers a ``click'' during interaction (\Cref{fig:existing}). In the current paradigm of designing and prototyping physical interfaces, we observed that the haptics of a physical input is often a by-product of the selected component (e.g. a push button, a knob); rather than an independent feature to consider. Furthermore, while the haptics of force feedback mechanisms can be characterized by their force-displacement (FD) curve; this curve is invisible during the design process and is usually empirically evaluated with physical measurements after fabrication \cite{liao_button_2020}.

\subsection{Shaping Passive Force Feedback}

Through this research, we developed \name, a new approach that enables designers---in particular industrial designers---to rapidly author and fabricate passive force feedback mechanisms for physical interfaces. \name\ mechanisms abstracts and simplifies passive force feedback systems into two-dimensional configurations (\Cref{fig:intro}A, B, C). These mechanisms leverage the material properties of POM (polyoxymethylene plastic) sheets and are simple to laser cut. They are composed of a compliant spring structure that engages with a moving slider during interaction (much like plastic snap fit buckles). By shaping the slider’s edge profile, designers can easily customize the haptic force feedback delivered by a \name\ mechanism---broadening and adding nuance to the haptics expressed by this class of mechanisms (\Cref{fig:intro}). To facilitate \name, we developed a computational design sandbox that enables designers to create and edit different force feedback mechanisms while providing real-time previews of their haptics by estimating their FD curves.

We discuss the development of \name\ in this paper and offer the following contributions to the design and fabrication of haptic physical interfaces:

\begin{enumerate}

    \item We present the \name\ approach for designing and fabricating passive and planar force feedback mechanisms with linear and rotary movements. These mechanisms are easy to fabricate with a laser cutter and they deliver a broad range of haptic force feedback profiles through simple customization.
    
    \item We share insights from our material exploration into designing and fabricating compliant structures with POM, as well as data from characterizing these compliant structures through virtual material simulations and real world measurements.
    
    \item We provide a computational design sandbox that enables designers to explore \name\ mechanisms while receiving real-time estimates of the corresponding FD curves. We discuss observations and insights about using this sandbox from pilot workshops with professional designers.
    
    \item We propose a series of interactive demonstrations with \name\ that showcase the potential applications of this approach for designing haptic physical interfaces.
    
\end{enumerate}
\section{Background \& Related Work}
\label{sec:relatedwork}

Mechanisms that provide interactive haptic feedback can be broadly categorized into \textit{active} and \textit{passive} devices: active devices apply external energy to control haptic feedback during interaction, while passive (or unpowered) devices only dissipate mechanical energy applied by humans during interaction \cite{hayward_haptic_2004}. In this research, we focus on \textit{passive} methods of delivering haptics. More specifically we investigate passive haptics through \textit{force feedback} \cite{van_oosterhout_ripple_2018, van_oosterhout_dynaknob_2019, swindells_role_2006}. In this section, we review the related work in HCI for designing passive force feedback mechanisms and highlight the important role of material investigation in such mechanisms. There is also a sizable body of research within HCI surrounding facilitating haptics design with authoring tools for active force feedback systems \cite{seifi_haptipedia_2019}. We discuss how these tools inspire \name\ and our development of exploration tools for passive haptic systems.

\subsection{Material-Driven Haptic Force Feedback}

Passive force feedback mechanisms essentially rely on the properties of physical materials to dissipate and redirect forces during interaction. For example, helical springs are structured to deliver reaction forces that oppose compression or tension, while rubberized brake pads generate friction that dampens movement. Beyond these classical mechanisms, HCI researchers have proposed many creative ways of leveraging physical materials for passive force feedback.

For example, researchers have demonstrated using magnets for passive force feedback in a tangible interface. These include plotting patterns of varying polarities on a magnetic sheet to deliver variable force feedback \cite{yasu_bump_2015, yasu_magnetact_2019, yasu_magnetic_2017}, embedding magnets in 3D printed parts to prototype haptic physical inputs \cite{zheng_mechamagnets_2018, zheng_mechamagnets_2019}, and creating haptic textures by dragging magnets across laser cut terrain \cite{wolf_haptic_2013}. In addition, researchers have demonstrated composing a haptic texture by configuring different magnets along a path and computing the resultant magnetic field \cite{ogata_magneto-haptics_2018}. Besides magnets, HCI researchers have also demonstrated passive haptic mechanisms using other materials such as paper \cite{chang_kirigami_2020} and 3D printing on fabric \cite{goudswaard_fabriclick_2020}.

The related work on physical materials and haptics reveal important trade-offs when designing passive haptic systems for physical interfaces. While passive material-driven haptic systems lack the on-the-fly reprogrammability of active haptic systems, they easily leverage digital fabrication processes such as 3D printing \cite{zheng_mechamagnets_2019, goudswaard_fabriclick_2020}, laser cutting \cite{chang_kirigami_2020}, and physical crafting processes \cite{yasu_magnetact_2019} to customize the haptic force feedback delivered. Material-driven haptic systems are thus appropriate in applications where the haptic feedback remains constant for a specific interaction. They are also better suited to explore haptic force feedback that is embedded into a more complicated physical interface; as compared to bulkier and more complex active haptic systems (such as \cite{ding_haply_2018, orta_martinez_evolution_2020}). In addition, exploring haptics via material-driven systems might be preferable or more approachable to certain communities (such as industrial designers); as opposed to exploring haptics via programming and electronics \cite{seifi_how_2020}.

We propose a new design approach for material-driven haptic systems. Previous material systems offer a discrete set of haptic building blocks (e.g. \cite{yasu_magnetact_2019, zheng_mechamagnets_2019, goudswaard_fabriclick_2020, chang_kirigami_2020}), or require 3D and multi-material assemblies (e.g. \cite{ogata_magneto-haptics_2018}).
We demonstrate how \name\ supports the authoring of arbitrary haptic force feedback profiles through shaping the 2D structure of laser cut POM sheets.

\subsubsection{Compliant Structures and Mechanisms}

Compliant structures are physical structures that contain elastic regions that flex in response to an applied force. Complete mechanisms can be constructed out of compliant structures and they offer certain advantages over movable joints; including a reduction in part count, simpler fabrication and assembly processes, and reduced wear and tear \cite{howell_compliant_2013}. HCI researchers have applied compliant structures to support flexible mounts and joints \cite{roumen_springfit_2019} or articulated mechanisms with laser cutting \cite{leen_lamifold_2020}, as well as sensing structures for 3D printed physical inputs \cite{savage_lamello_2015} or embedded springs in 3D printed parts \cite{he_ondulxe9_2019}. We detail our findings on designing and characterizing two types of compliant springs with laser cut POM that exhibits almost linear force-deflection relationships.

\subsection{Quantifying Haptic Force Feedback}

Our perception of haptics during interaction is affected by a complex variety of factors, including cutaneous stimulation (e.g. mechanical vibrations on the skin) and kinesthetic forces (e.g. the weight of an object) \cite{erp_setting_2010}. HCI researchers have proposed a few methods of quantifying the latter. Among these methods, the FD curve is often used as a preliminary way of considering force feedback delivered by a mechanism that displaces with one degree of freedom (e.g. buttons and knobs) \cite{liao_button_2020, kim_haptic_2013, doerrer_simulating_2002, van_oosterhout_facilitating_2020, van_oosterhout_dynaknob_2019}. It is important to note that the FD curve does not comprehensively communicate all the haptic nuances of interacting with a physical input. For instance, static FD calculations do not capture changes due to different interaction velocities and they do not account for vibrations (cutaneous stimulation) that participate in the haptics \cite{liao_button_2020}. However, for designers unfamiliar with haptics, FD curves offer an approachable way to consider the haptics of physical inputs; such as the ``click'' of a button \cite{chang_kirigami_2020, kim_haptic_2013}, or the stops along a knob’s rotation \cite{van_oosterhout_knobology_2018}. With this in mind, we adopt the FD curve in this work to approximate and communicate the haptics delivered by \name\ mechanisms.

\subsection{Authoring Haptic Force Feedback}

HCI researchers have developed tools for authoring haptic force feedback mechanisms supported by FD curves. Button simulators such as \cite{liao_one_2018, liao_button_2020, doerrer_simulating_2002} offer a physical button on a stage with active force feedback control. Different force displacement curves can be loaded into the simulator which then renders the corresponding haptic force feedback when the button is pressed. For rotational movements, \textit{Feelix} enables designers to explore and edit the haptic force feedback of motor-driven rotary devices through their FD (angular displacement) curves \cite{van_oosterhout_facilitating_2020}. We are inspired by these prior work and extend the concepts they present to passive force feedback mechanisms.

Beside the specific use of FD curves to facilitate haptic force feedback design, there are also broader guidelines to consider when developing a haptic authoring tool. \cite{van_oosterhout_facilitating_2020, seifi_how_2020} succinctly captures and categorizes these guidelines, and we actively considered and incorporated these recommendations into the computational design sandbox developed for \name. These include supporting an iterative design process in exploring and refining haptic mechanisms \cite{seifi_how_2020, swindells_role_2006}, simple set up \cite{schneider_tactile_2015}, real-time visual previews to support debugging \cite{seifi_how_2020, schneider_tactile_2015, seifi_vibviz_2015}, as well as flexibility in software use \cite{seifi_how_2020}.
\section{\name: Working Principle}
\label{sec:workingprinciple}

\begin{figure}[h]
    \centering
    \includegraphics[width=\linewidth]{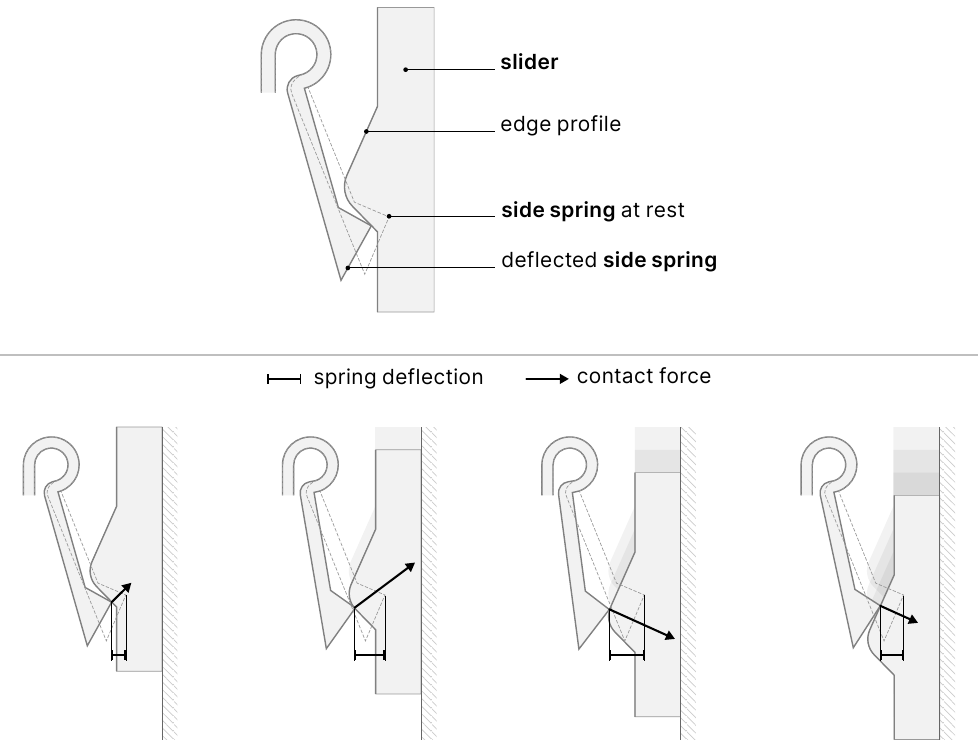}
    \Description[Shape-Haptics working principle]{Shape-Haptics mechanisms comprises two main parts, a compliant side spring that moves along a rigid sliding edge profile. Force feedback depends on the magnitude of spring deflection and direction of spring force.}
    \caption{\name\ working principle: the slider's edge profile engages with the side spring as it moves and delivers varying force feedback.}
    \label{fig:principle}
\end{figure}

\begin{figure*}[tbp]
    \centering
    \includegraphics[width=0.9\textwidth]{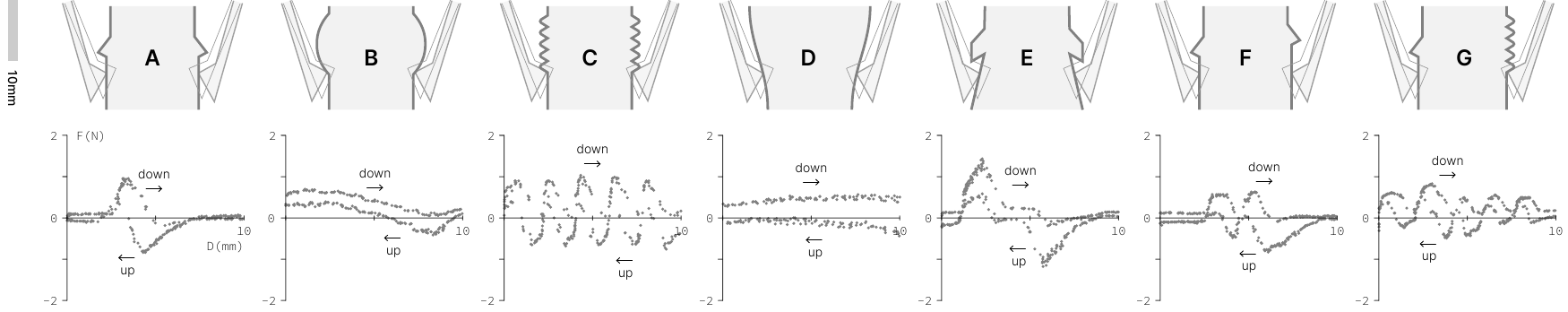}
    \Description[Seven Shape-Haptics mechanism variations]{Seven Shape-Haptics mechanism variations and their force displacement curves captured by a physical measurement rig.}
    \caption{Double-sided profile test. The force-displacement data was captured via a FD measurement rig.}
    \label{fig:seventest}
\end{figure*}

\name\ draws inspiration from mechanisms used in existing physical inputs for passive force feedback. For example, mechanical key switches ``click'' when a spring brushes against a bump, while knobs ``tick'' when the spring moves between detents as it rotates (\Cref{fig:existing}). We extend the working principle of these ubiquitous objects with \name\ and demonstrate the broader and more nuanced range of passive force feedback possible with this class of mechanisms. \name\ mechanisms rely on two key components: (A) a side spring in the form of a compliant structure and (B) a rigid slider with an edge profile that moves across the side spring (\Cref{fig:principle}). \name\ mechanisms deliver variable force feedback along its displacement by changing the side spring's compression (magnitude of force) as well as the contact angle between the side spring and edge profile (direction of force). In addition, the friction encountered by the sliding component affects the overall force feedback.

\begin{figure}[h]
    \centering
    \includegraphics[width=\linewidth]{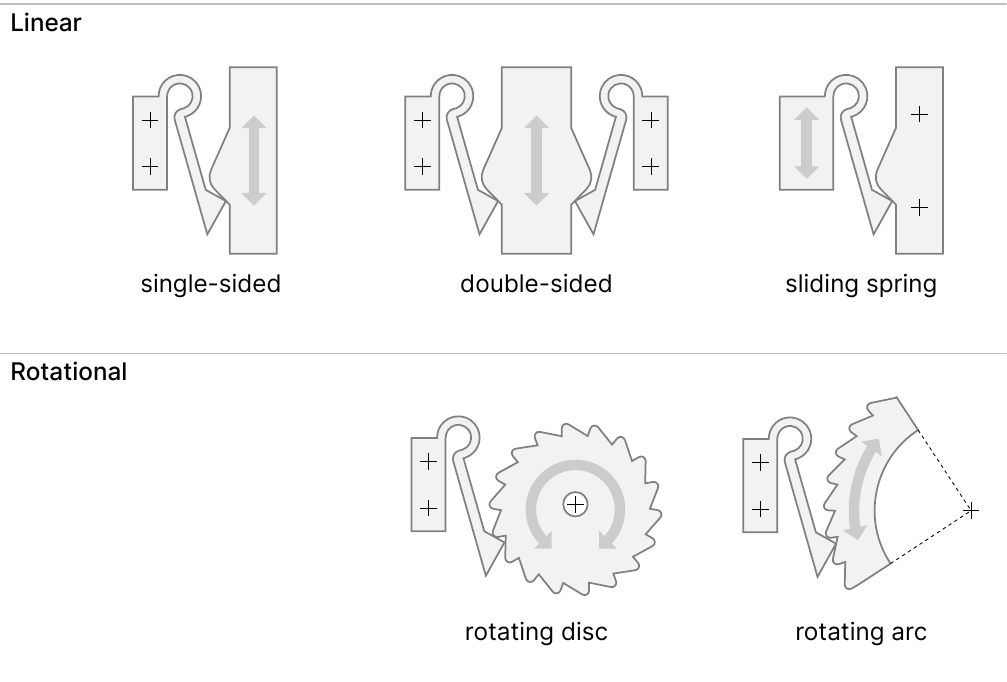}
    \Description[Shape-Haptics fundamental configurations]{Shape-Haptics has five fundamental configurations, organized into two categories: linear movements and rotational movements. For linear movements, mechanisms can be single-sided, double-sided, and with a sliding spring (instead of a sliding profile). For rotational movements, mechanisms can be a rotating disc or a rotating arc.}
    \caption{\name\ fundamental configurations.}
    \label{fig:config}
\end{figure}

\subsection{Configurations}

\name\ mechanisms are constrained to move along one degree of freedom across its plane. With this in mind, we developed two fundamental configurations (linear and rotation), as well as variations within each configuration (\Cref{fig:config}). Furthermore, the planar structure of \name\ mechanisms enables two profiles to be incorporated within one slider. These double-sided profiles can be symmetrical to increase the overall magnitude of the force feedback; or, they can be asymmetrical to provide more detailed force feedback by adding the different forces that each profile delivers (for examples, refer to \Cref{fig:seventest}F, G). For instance, designers might use an asymmetrical double-sided profile configuration to circumvent the fabrication resolution of the laser cutter, or to explore a more nuanced overall FD curve by manipulating two simpler profiles (rather than designing one complicated profile). While we believe that there is potential for more complicated configurations by extending the working principle (such as by stacking different planar movements to support movements in multiple degrees of freedom), we focus on the fundamental configurations and their variations in this paper.

\begin{figure}[h]
    \centering
    \includegraphics[width=\linewidth]{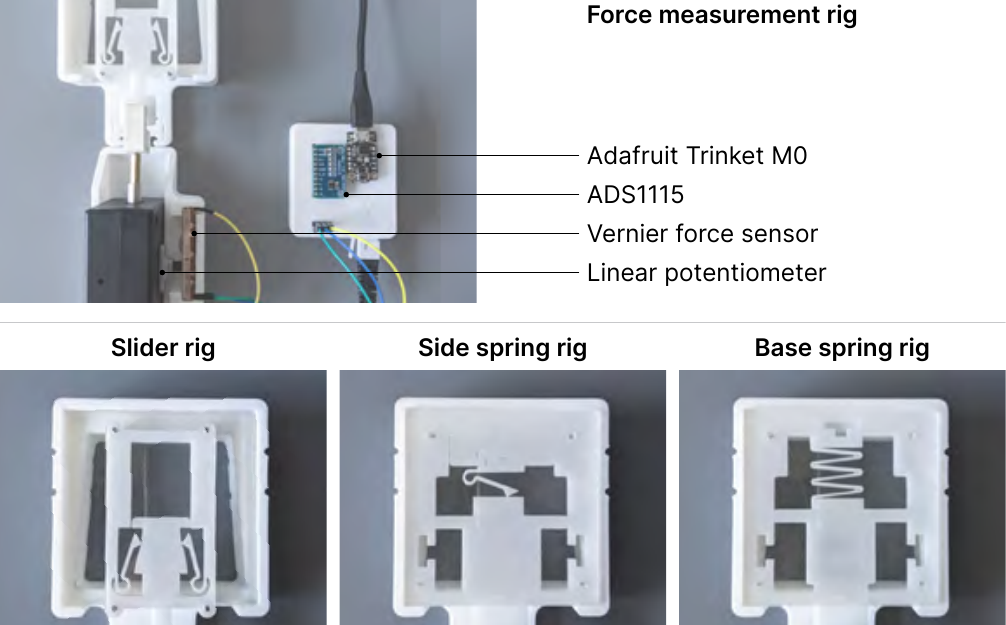}
    \Description[Force-displacement measurement rig]{The force-displacement measurement rig comprises a Vernier force sensor coupled to a linear potentiometer. These sensors are read by a 16-bit analog to digital converter and the Adafruit Trinket M0 microcontroller. We use this rig to measure different mechanisms, including the side springs, base springs, and whole Shape-Haptics mechanisms.}
    \caption{FD measurement rig. Force (N) is captured by a Vernier force sensor while displacement (mm) is captured by a linear sliding potentiometer.}
    \label{fig:rig}
\end{figure}

Figure \ref{fig:seventest} shows different double-sided slider profiles with a total displacement of 10mm moving against the same side springs. We fabricated these mechanisms and measured their corresponding FD curves with a custom force sensing rig (\Cref{fig:rig}). This rig consists of a Vernier force sensor coupled to a linear sliding potentiometer. These sensors were connected to a 16-bit analog-to-digital converter (ADS1115) read by a microcontroller (Adafruit Trinket M0). We laser cut and tested three samples for each profile. As demonstrated by these examples, editing the contours of the sliding profile changes the FD curves that these mechanisms deliver. We also demonstrate two asymmetrical double-sided profiles that stagger the same profile (\Cref{fig:seventest} F vs. A) and mix two different profiles (\Cref{fig:seventest} A+C=G).

\section{Material Investigation: Laser Cutting POM Mechanisms}
\label{sec:material}

\name\ makes use of flat and compliant spring structures to deliver force feedback during interaction. Such compliant structures enable us to reduce the number of parts in the mechanism and thus simplify its assembly, as well as construct the entire mechanism from one material. Compliant structures should be fabricated with materials that have a high ratio of yield strength to tensile modulus \cite{howell_compliant_2013}---meaning it should be able to withstand high loads while remaining flexible. We were interested in facilitating the fabrication of these mechanisms with the laser cutter as it is a personal fabrication tool found in many maker spaces and design studios. However, common laser cut materials are not ideal for such compliant mechanisms, particularly at the scale of \name\ mechanisms (1.0--2.5mm wall thickness). For example, acrylic has significantly reduced tensile strength after laser cutting (\textasciitilde35MPa \cite{nigrovic_comparison_2017}) while plywood sheets exhibit anisotropic material strength depending on the direction of wood fibers.

We were inspired by \cite{amymakesstuff_flexure-fun_2021} and their work on laser cut compliant mechanisms with POM. POM (trade names include Delrin, Sustarin) is an engineering thermoplastic with high tensile strength (\textasciitilde67MPa \cite{rochling_sustarin_nodate}), low friction, and low wear, making it suitable for high impact industrial and mechanical applications. For this research, we used 3mm thick Sustarin sheets \cite{rochling_sustarin_nodate} found at a local supplier and conducted our investigations on this specific brand of POM.

\subsection{Laser Cutting POM}

\begin{figure}[h]
    \centering
    \includegraphics[width=\linewidth]{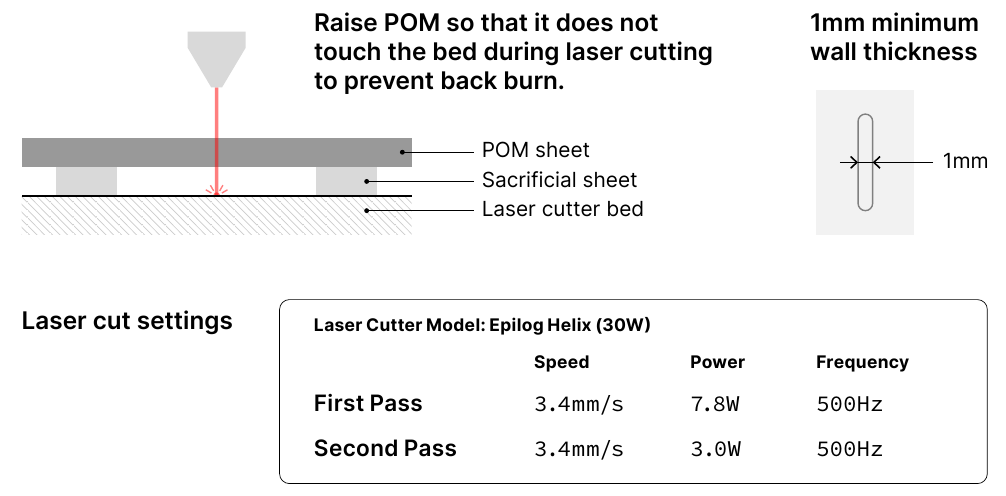}
    \Description[Laser cutting guidelines for 3mm POM sheets]{Laser cutting guidelines for 3mm POM sheets. Raise POM sheet from the laser cutter bed with sacrificial material to prevent back burn. Recommended minimum wall thickness is 1mm. We used a 30W Epilog Helix laser cutter. To cut 3mm POM sheets, we used speed 3.4mm/s, power 7.8W, and frequency 500Hz for the first pass; and speed 3.4mm/s, power 3.0W, and frequency 500Hz for the second pass.}
    \caption{Guidelines for laser cutting POM sheets.}
    \label{fig:laser}
\end{figure}

\begin{figure*}[tbp]
    \centering
    \includegraphics[width=\textwidth]{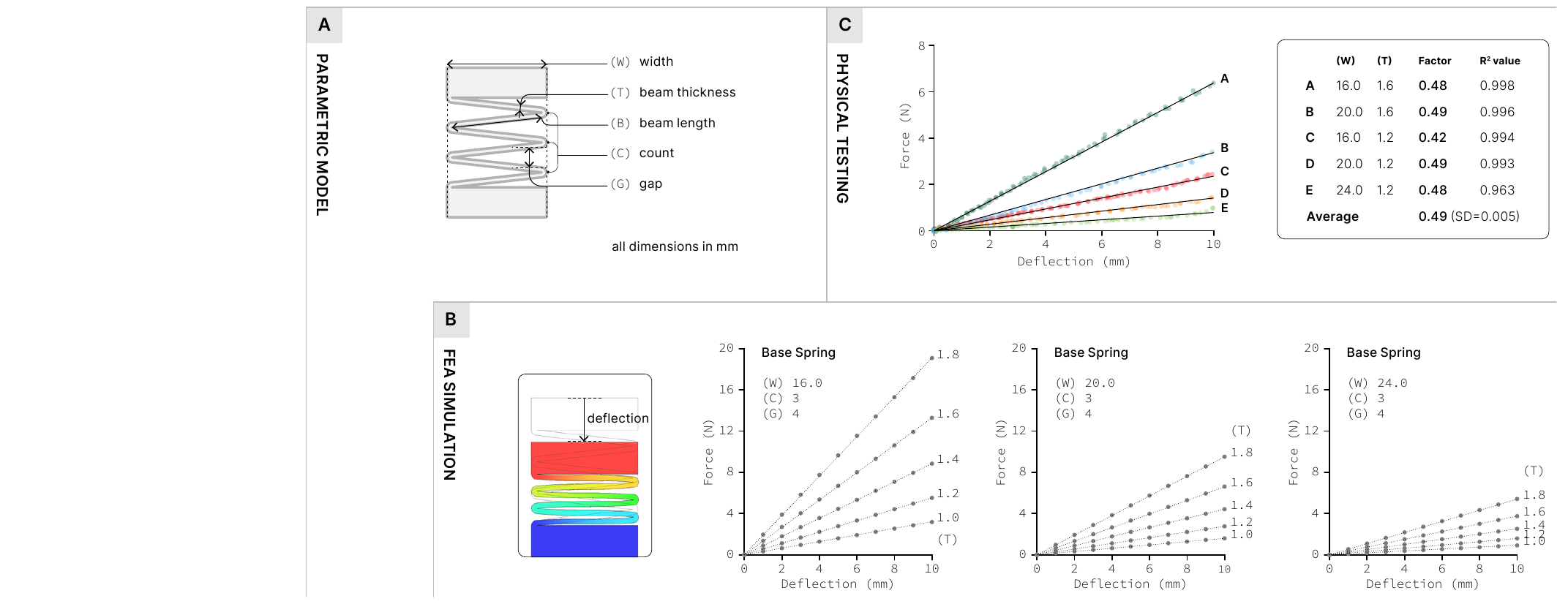}
    \Description[Characterization of base spring]{The base spring was abstracted into a parametric model with five variables. Virtual finite element analysis simulations were conducted on base spring variations to derive their force-deflection slope. A subset of these base springs were fabricated and the physical force-deflections slope measured and compared to the simulated results.}
    \caption{A: Parametric base spring model. B: FEA simulation results and best fit lines. C: Comparing and adjusting physical measurements to simulations results.}
    \label{fig:basechar}
    \vspace*{12pt}
\end{figure*}

Laser cutting POM sheets was more challenging than we anticipated; especially for small intricate cutting patterns found in compliant structures (such as in \Cref{fig:sidechar}, \ref{fig:basechar}). The fabricated parts have a tendency to fuse back to the sheet when the laser exits the material. In addition, thin walls might warp if they are too close or cut in quick succession. From our material exploration, we arrived at a set of fabrication guidelines that we used for all subsequent laser cut POM parts. These guidelines are captured in \Cref{fig:laser}.

\subsection{Side Spring Development and Testing}

The side spring is a key component in \name\ mechanisms that interacts with the slider to deliver variable force feedback. We took three design criteria into consideration when exploring compliant geometries for this spring: (1) The side spring should be as compact as possible relative to the rest of the mechanism. (2) The force it exerts should be predictably proportional to its deflection (ideally linear) to facilitate computational design. (3) The spring should deflect fully without warping.

\begin{figure}[h]
    \centering
    \includegraphics[width=\linewidth]{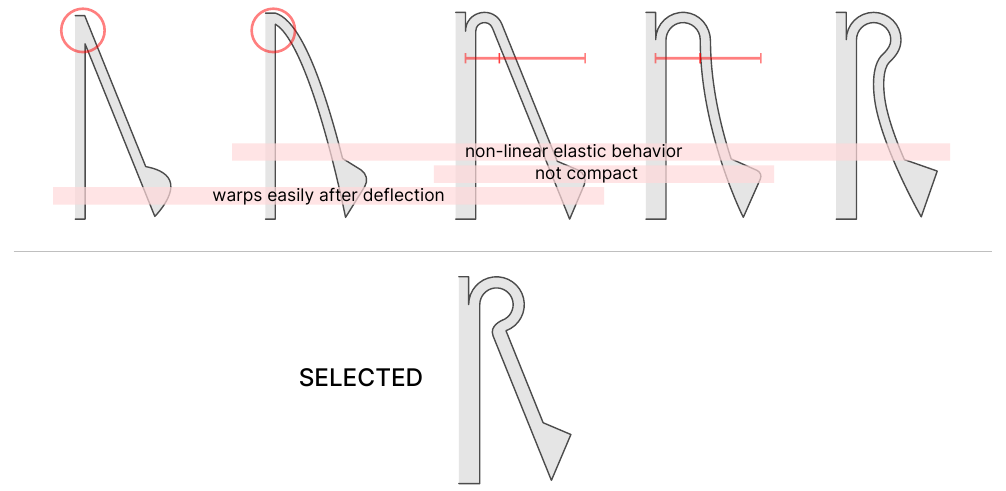}
    \Description[Side spring geometry exploration]{Various side spring geometries were explored. The selected side spring exhibited close to linear elastic behavior, was compact in terms of size change between compressed and uncompressed states, and did not warp easily after deflection.}
    \caption{Side spring geometry exploration.}
    \label{fig:springexplore}
\end{figure}

\begin{figure*}[tbp]
    \centering
    \includegraphics[width=\textwidth]{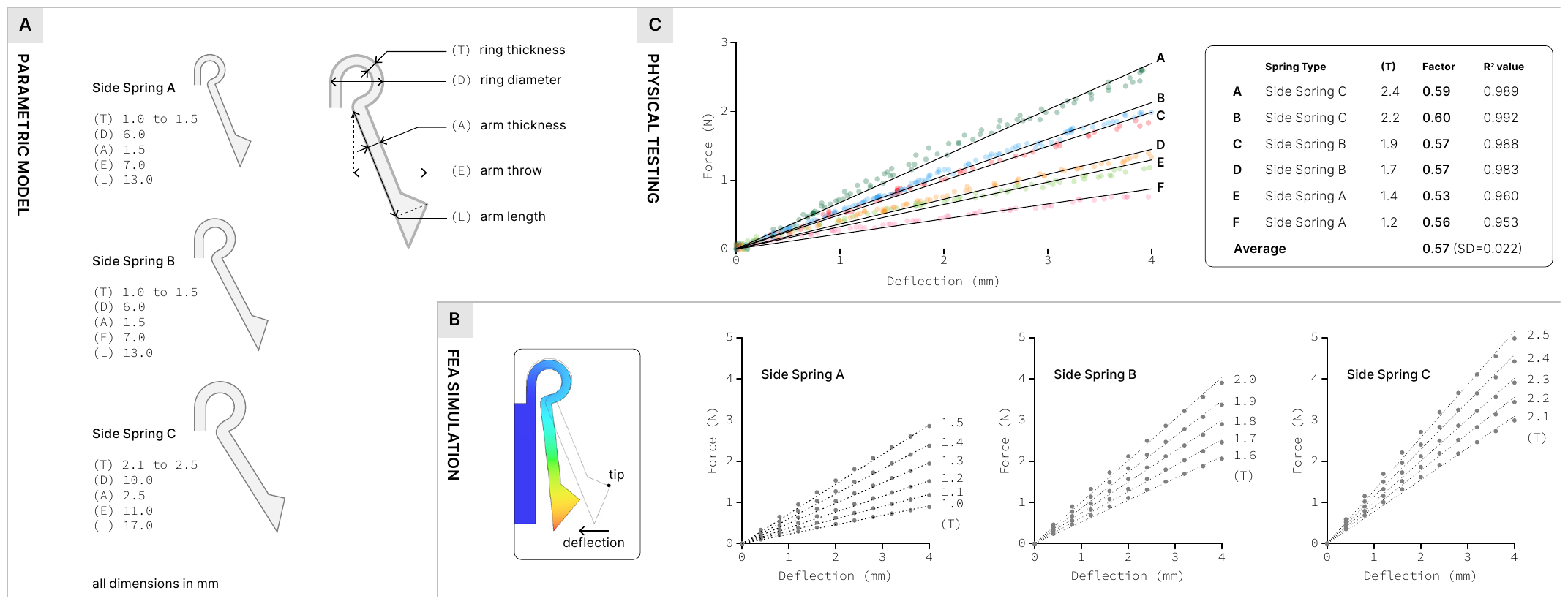}
    \Description[Characterization of side spring]{The side spring was abstracted into a parametric model with five variables. Three versions of the side spring was generated from this parametric model, each with different ring thicknesses. Virtual finite element analysis simulations were conducted on these side spring variations to derive their force-deflection slope. A subset of these side springs were fabricated and the physical force-deflections slope measured and compared to the simulated results.}
    \caption{A: Parametric side spring model and the three variations we use for Shape-Haptics. B: FEA simulation results and best fit lines. C: Comparing and adjusting physical measurements to simulations results.}
    \label{fig:sidechar}
\end{figure*}

Figure \ref{fig:springexplore} illustrates the different spring geometries we explored, along with annotations that indicate their different limitations. The final geometry we used behaves like a torsion spring. This spring’s deflection occurs along the ring while the straight arm remains rigid (for arm thickness > ring thickness). This spring also demonstrates an almost linear relationship between force and deflection (\Cref{fig:sidechar}). The tip of the spring was shaped with 45 degree slopes on either side to facilitate sliding across a rigid edge profile in both directions. We developed a parametric model for this side spring geometry and generated three versions of this side spring (\Cref{fig:sidechar}A) that offer different spring coefficients for force feedback. These side springs have a maximum deflection of 4mm and apply a maximum contact force between approximately 0.75N to 3N (\Cref{fig:sidechar}C).

\subsection{Calculating Side Spring Coefficient}

We conducted FEA simulations as well as physical measurements to verify the spring's force-deflection behavior and quantify its spring coefficient. We used this data to support our development of the computational design sandbox for \name\ mechanisms (\Cref{sec:sandbox}).

We used a non-linear static stress study in Autodesk Fusion 360 to simulate deflecting the side spring and measured the reaction force at its tip (\Cref{fig:sidechar}B). We used the material specifications of POM detailed in \cite{rochling_sustarin_nodate} and performed 16 studies across different ring thickness values for side spring A, B, and C (\Cref{fig:sidechar}A). Measurements were taken at regular intervals along the spring’s displacement. \Cref{fig:sidechar}B displays the results from these simulations. A line of best fit was calculated for each study (with the intercept set to 0). \Cref{tab:sideslope} contains the slope of the best-fit lines (the spring coefficient). These best fit lines all had $R^2$ values more than 0.99.

These simulations provide insight into the relationship between the different geometric parameters of the side springs and their spring coefficient. However, they do not account for the loss of material integrity due to laser cutting \cite{nigrovic_comparison_2017}. We observed that the laser cut springs were weaker than the data reported from the FEA simulations. We therefore investigated the difference between simulated results and real-world measurements. We took a subset of 6 side springs variations (out of the 16 we studied through FEA simulation) and conducted real world force-deflection measurements on these springs. We used the same rig (\Cref{fig:rig}) and measured three samples for each spring variation. With the physical measurements, we adjusted the simulation data with a multiplication factor to find the best fit for each physical test and its corresponding simulation data (\Cref{fig:sidechar}C). From this empirical study, we found an average multiplication factor of 0.57 (SD = 0.022) between the simulated data and physical measurements. We multiplied this factor to the slopes from the FEA simulation to derive a lookup table of spring coefficients for the side spring model (\Cref{tab:sideslope}).

\subsection{Base Spring Development and Testing}

Besides the side spring, we also developed a secondary base spring for linear \name\ mechanisms. These base springs return the slider to its rest position (e.g. \Cref{fig:intro}A). We adopted the same criteria and process as the side spring to develop this base spring. We began by exploring different geometries and arrived at the design illustrated in \Cref{fig:basechar}A. Essentially, this base spring behaves like a series of cantilevered beams that flex slightly under compression. Following the standard beam deflection formula, the arm length, arm thickness, and material thickness are variables that influence the spring coefficient. We then developed a parametric model of the base spring (\Cref{fig:basechar}A) with these key\vadjust{\pagebreak} parameters and ran FEA simulations on 15 variations of the spring (\Cref{fig:basechar}B). Best fit lines were applied to this data to derive the slope/spring coefficient (the data was linear, $R^2=1$ for all variations tested). We also took physical measurements of 5 base spring variations and compared it to the corresponding simulation data. From this comparison, we found an average multiplication factor of 0.49 (SD = 0.005) between the physical measurements and FEA data. This is slightly lower than the side spring factor (0.57) and we posit that it can be attributed to the more intricate base spring design, which exposes the material to a longer fabrication time in the laser cutter. Similarly, we multiplied this factor to the slopes from the FEA simulation to derive a lookup table of spring coefficients for the base spring model (\Cref{tab:baseslope}).

\begin{table}[h]
\small
\centering
\begin{tabular}{C{0.5in} C{0.15in} C{1.0in} C{1.0in}} 
  \toprule
  \textit{Side Spring} & \textit{(T)} & \textit{FEA Slope (N/mm)} & \textit{Adjusted Slope (0.57x)}
  \\
  \arrayrulecolor{black!30}\midrule
  \textbf{A} 
  & 1.0 & 0.23 & 0.13   \\
  & 1.1 & 0.30 & 0.17   \\
  & 1.2 & 0.39 & 0.22   \\
  & 1.3 & 0.50 & 0.28   \\
  & 1.4 & 0.61 & 0.35   \\
  & 1.5 & 0.74 & 0.42   \\
  \arrayrulecolor{black!30}\midrule
  \textbf{B} 
  & 1.6 & 0.53 & 0.30   \\
  & 1.7 & 0.64 & 0.36   \\
  & 1.8 & 0.75 & 0.43   \\
  & 1.9 & 0.87 & 0.50   \\
  & 2.0 & 1.01 & 0.58   \\
  \arrayrulecolor{black!30}\midrule
  \textbf{C} 
  & 2.1 & 0.77 & 0.44   \\
  & 2.2 & 0.89 & 0.51   \\
  & 2.3 & 1.01 & 0.58   \\
  & 2.4 & 1.15 & 0.65   \\
  & 2.5 & 1.29 & 0.74   \\
  
  \bottomrule
\end{tabular}

\Description[Side spring coefficient look up table]{A look up table with 16 side spring coefficients based on the adjusted slope based on simulated results and physical measurements.}
\caption{Slopes (spring coefficients) derived from FEA simulations of different side spring variations and their adjusted values after physical measurements.}
  \label{tab:sideslope}
  
\end{table}

\vspace{-4mm}

\begin{table}[h]
\small
\centering
\begin{tabular}{C{0.2in} C{0.2in} C{0.3in} C{0.8in} C{1.0in}} 
  \toprule
  \textit{(W)} & \textit{(T)} & \textit{(B)} & \textit{FEA Slope (N/mm)} & \textit{Adjusted Slope (0.49x)}
  \\
  \arrayrulecolor{black!30}\midrule
  16.0 
  & 1.0 & 14.14 & 0.32 & 0.16   \\
  & 1.2 & 13.75 & 0.55 & 0.27   \\
  & 1.4 & 13.35 & 0.89 & 0.43   \\
  & 1.6 & 12.96 & 1.33 & 0.65   \\
  & 1.8 & 12.56 & 1.91 & 0.94   \\
  
  \arrayrulecolor{black!30}\midrule
  20.0 
  & 1.0 & 18.11 & 0.16 & 0.08   \\
  & 1.2 & 17.71 & 0.28 & 0.14   \\
  & 1.4 & 17.32 & 0.44 & 0.22   \\
  & 1.6 & 16.92 & 0.66 & 0.32   \\
  & 1.8 & 16.52 & 0.95 & 0.47   \\
  
  \arrayrulecolor{black!30}\midrule
  24.0 
  & 1.0 & 22.09 & 0.09 & 0.04   \\
  & 1.2 & 21.69 & 0.16 & 0.08   \\
  & 1.4 & 21.29 & 0.25 & 0.12   \\
  & 1.6 & 20.90 & 0.38 & 0.18   \\
  & 1.8 & 20.50 & 0.54 & 0.27   \\

  \bottomrule
\end{tabular}
\Description[Base spring coefficient look up table]{A look up table with 15 base spring coefficients based on the adjusted slope based on simulated results and physical measurements.}
\caption{Slopes (spring coefficients) derived from FEA simulations of different base spring variations and their adjusted values after physical measurements.}
  \label{tab:baseslope}
  
\end{table}

\vspace{5mm}

\begin{figure}[h]
    \centering
    \includegraphics[width=\linewidth]{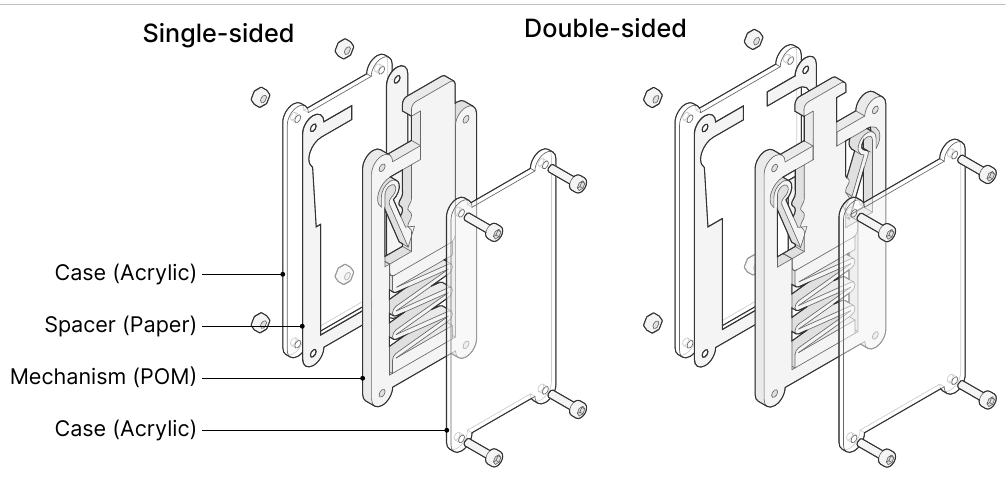}
    \Description[Shape-Haptics mechanism swatches exploded view]{Exploded view of the single sided and double sided linear Shape-Haptics mechanism swatches that the computational design sandbox supports.}
    \caption{Exploded views for single and double sided \name\ mechanisms supported by the sandbox.}
    \label{fig:exploded}
\end{figure}

\section{Developing a Computational Design Sandbox}
\label{sec:sandbox}

Equipped with the material and fabrication characteristics we uncovered, we built a computational design sandbox for exploring and building \name\ mechanisms\footnote{This sandbox was developed as a client-side web application and used the \textit{Paper.js} library for 2D vector drawing and calculations. It can be accessed at \textit{\url{https://interactive-materials.github.io/shape-haptics}}.}. The sandbox supports the design and fabrication of hand-held \name\ swatches that move linearly (\Cref{fig:exploded}). Designers can opt to design a single-sided mechanism (one edge profile on the slider) or a double-sided mechanism. These mechanisms are designed to be rapidly fabricated, assembled, and disassembled to facilitate design exploration.

\subsection{User Interface Design}

\begin{figure*}[tbp]
    \centering
    \includegraphics[width=\textwidth]{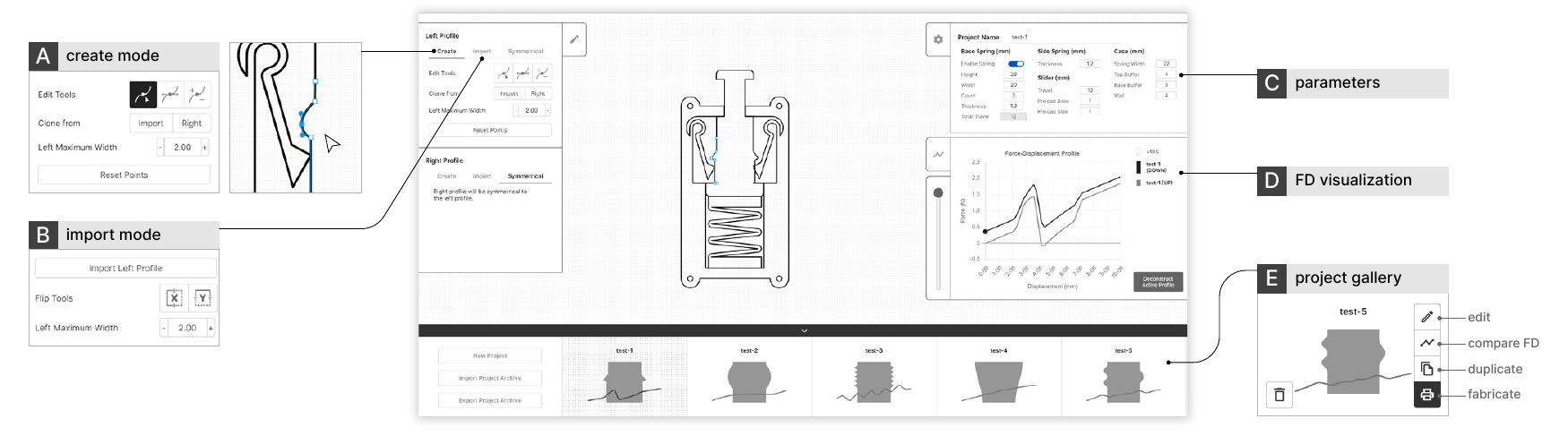}
    \Description[Sandbox user interface]{The sandbox user interface is organized into four main areas. A canvas in the center where the mechanism is edited. A left panel with different editing options. A right panel with customizable spring parameters and the force-displacement curve visualization. A bottom panel with a gallery of different projects.}
    \caption{Sandbox user interface.}
    \label{fig:ui}
\end{figure*}

\textbf{Edit Panel:} Designers edit the sliding profile via two modes: \textit{create} (\Cref{fig:ui}A) and \textit{import} (\Cref{fig:ui}B). In \textit{create} mode, profiles are displayed as a polyline/spline defined by a series of segment points; similar to vector editing tools like Adobe Illustrator. Segment points can be added, removed, moved, and pulled to change the profile. In \textit{import} mode, profile designs in the form of SVG files can be loaded into the sandbox. This enables designers to use a digital design tool of their choice to construct these profiles. For double-sided projects, there is an additional \textit{symmetrical} mode to mirror the opposite profile.

\textbf{Parameters Panel:} This panel (\Cref{fig:ui}C) enables designers to modify the geometric parameters of the mechanism. This includes important spring dimensions like \textit{thickness} and \textit{width} which affect the spring coefficient, as well as \textit{travel} which affects the maximum displacement of the slider.

\textbf{FD Visualization Panel:} This panel (\Cref{fig:ui}D) displays the estimated FD curves of the active project. There is also an option to toggle the visualization to display separated FD curves for the left and right profile in the case of an asymmetrical double-sided project. We elaborate on the FD visualization in the following section.

\textbf{Gallery:} Multiple projects can be created in one session and these projects are stored in the gallery (\Cref{fig:ui}E). Gallery projects can be duplicated and worked on separately; such as when iterating and making small adjustments to a particular profile. The fabrication drawings for a project can be exported from the gallery in the form of an SVG file for laser cutting. Designers can also save the entire gallery as a project archive file---and load it back into the sandbox in the future.

\begin{figure*}[tbp]
    \centering
    \includegraphics[width=0.95\textwidth]{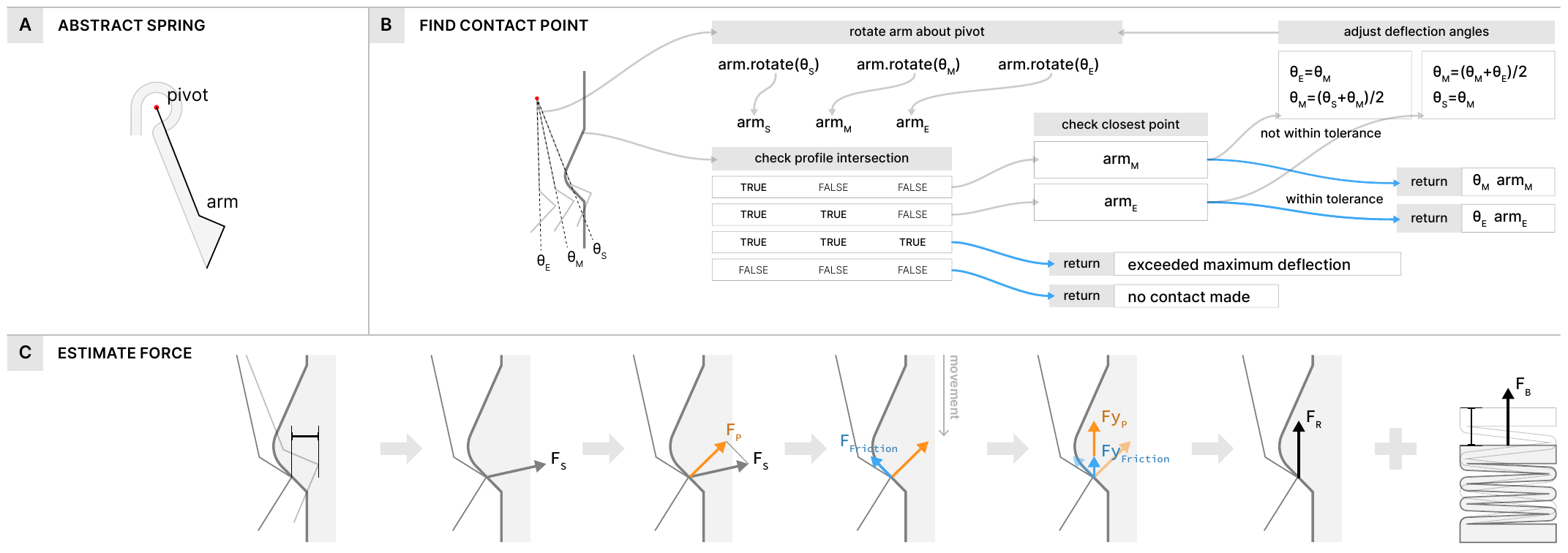}
    \Description[Algorithm for calculating force feedback]{The side spring is abstracted into a simple arm profile that rotates about a pivot. The contact point between the profile and simplified arm model is computed by recursive rotating the arm till their closest points are within tolerance without intersecting. The force feedback is computed by resolving vertical components of the spring forces and friction.}
    \caption{A: Abstracted side spring model comprising an arm that rotates about a pivot point. B: Flow diagram for algorithm to find contact point between side spring and sliding profile. C: Flow diagram for estimating reaction force at a particular displacement.}
    \label{fig:algo}
\end{figure*}

\subsection{Estimating Force Feedback}

We developed an algorithm that estimates the FD curve of a \name\ mechanism. This algorithm works in two phases: (1) finding the point of contact between the side spring and the slider’s profile at a given displacement, followed by (2) estimating the forward and reverse force feedback generated.

To find the point of contact, we first abstract the side spring into a simplified arm model (\Cref{fig:algo}A). The edge profile of the slider is isolated and shifted to the specified displacement. The algorithm then employs a recursive approach of halving the deflection angle of the arm, starting with the full deflection region of the spring. The candidate that is closest to the profile without intersecting it is shortlisted, and the process is repeated till the closest point is within a specified distance from the profile (we use 0.005mm in the sandbox). The algorithm then returns the deflection angle of the arm and closest point on the profile for the next step. \Cref{fig:algo}B illustrates this phase.

With the contact point established, the next phase of the algorithm calculates the normal and tangent vectors on the profile at the point of contact. It then retrieves the spring coefficients from the lookup table (\Cref{tab:sideslope}, \ref{tab:baseslope}) and follows the steps below to estimate the forward and reverse force feedback at the specified displacement. This estimation employs a static free body diagram of a linear mechanism as illustrated by \Cref{fig:algo}C. The force feedback is indicated by $F_R$ (the reaction force at the top of the slider).

\begin{enumerate}
    \item Calculate $F_S$ (spring force) by taking the product of the side spring’s deflection and the spring coefficient.
    
    \item Project $F_S$ onto the normal of the contact point along the sliding profile to get $F_P$ (the force acting on profile).
    
    \item Calculate $F_{Friction}$: product of $F_P$ and material’s friction coefficient. We used POM’s dynamic friction coefficient value of 0.21 \cite{rochling_sustarin_nodate}.
    
    \item Sum the vertical components of $F_P$ and $F_{Friction}$ to get $F_R$. When the movement of the slider is reversed, friction acts in the opposite direction. This results in a different $F_R$.
    
    \item Calculate the base spring force $F_B$ (product of the base spring coefficient with the magnitude of displacement) and add it to $F_R$.

\end{enumerate}

By calculating $F_R$ at regular displacement intervals, we can interpolate the points and estimate the FD curve for a given slider profile. $F_R=0$ if no contact is made and a warning is issued if the contact point exceeds the side spring's maximum deflection.

\begin{figure*}[tbp]
    \centering
    \includegraphics[width=0.9\textwidth]{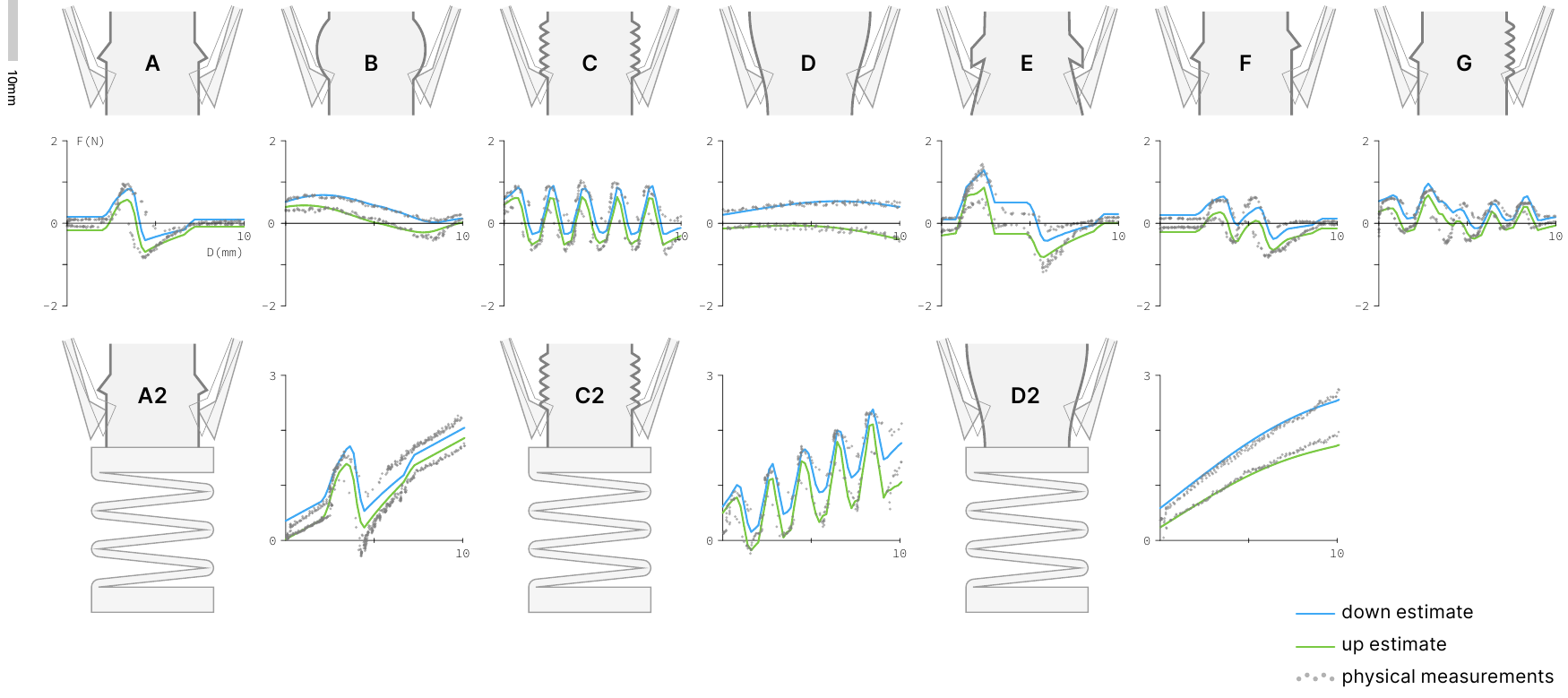}
    \Description[Comparing estimated force displacement curves to actual measurements]{The force displacement curves estimated by the sandbox for 7 different Shape-Haptics mechanisms were compared against actual physical measurements.}
    \caption{Comparison of FD estimations to physical measurements. Top, A--G: Comparing estimated FD curve to physical measurements (\Cref{fig:seventest}). Bottom A2--C2: Base spring included in both physical measurements and estimation.}
    \label{fig:compare}
\end{figure*}

\subsection{Visualizing and Interacting with the FD Curve}

The force feedback algorithm runs as a co-routine in the sandbox and iteratively estimates the force feedback at different displacements across the full travel distance of the slider. The arrays of FD values for both forward and reverse directions are interpolated into two FD curves and displayed as a line chart in the sandbox. These curves are updated whenever the profiles are edited to provide designers with real time estimates of the resulting haptic force feedback.

\begin{figure}[h]
    \centering
    \includegraphics[width=\linewidth]{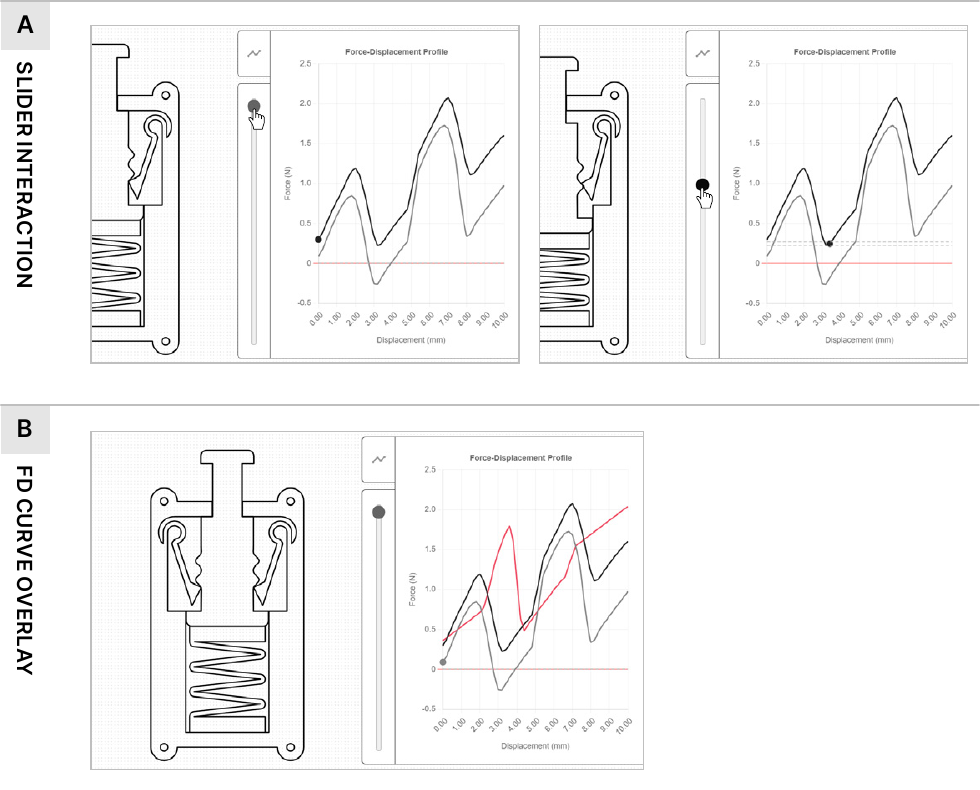}
    \Description[Virtual interactions with Shape-Haptics mechanisms in the sandbox]{Designers can virtually interact with Shape-Haptics mechanisms in the sandbox by pulling a range slider. As the range slider is pulled, the mechanism moves to show the contact point between profile and side spring. The corresponding point along the estimated force-displacement curve is also indicated. Estimated force-displacement curves from other projects can be overlaid on top of the active profile for comparison.}
    \caption{A: Range slider interaction, moving the slider moves the mechanism and highlights the corresponding point in the FD visualization. B: Overlaying FD curves for comparison.}
    \label{fig:slider}
\end{figure}

We added a range slider in the UI to give this information more interactivity (\Cref{fig:slider}A). Pulling the range slider will move the mechanism accordingly in the sandbox and displays how the sliding profile interacts with the springs at a specific displacement. The corresponding point on the FD curve is also highlighted. The FD curves of other projects in the gallery can also be selected and overlaid onto the curves for the active project (\Cref{fig:slider}B). This enables designers to compare the haptic force feedback of different mechanisms via the estimated FD data; for instance, when incrementally editing a certain profile shape. 

\subsection{Comparing Estimated FD Curves and Physical Measurements}

We compared the estimated FD curves generated by the sandbox (\Cref{fig:compare}) to the same seven profiles and their physical FD measurements (\Cref{fig:seventest}) to evaluate its predictive capability. In addition, we compared three of the profiles with the base spring included (\Cref{fig:compare} A2, B2, C2). We overlaid the estimated FD curves for downwards and upwards movements onto data from the physical measurements.

Qualitatively, the estimated FD curves matched the overall shape of the physical measurements. Peaks and valleys lined up between both sets of data and the estimated curves captured the same trends as the physical measurements (i.e. how the gradient of the curve varies over displacement). The two main discrepancies we observed can be described as gradient differences between an estimated slope and the measured slope, as well as parallel offsets in measured values. We hypothesize that these differences are caused by inaccuracies due to fabrication and assembly tolerances; such as an inconsistent laser cut kerf and the slider shifting horizontally while moving vertically during physical measurements. These differences might also be due to dynamics that our estimation method cannot account for (such as the considerations put forth by \cite{liao_button_2020} and frictional inconsistencies due to material and fabrication variations).

We found the estimated FD curves helpful in providing a first layer of information about the force feedback delivered by \name\ mechanisms during the digital design process; before committing to fabricating the part. Throughout this research, we used the sandbox to explore and get a sense of the haptics offered by different profiles, especially when we were developing the different application examples (\Cref{sec:propositions}).

\begin{figure*}[h]
    \centering
    \includegraphics[width=0.9\textwidth]{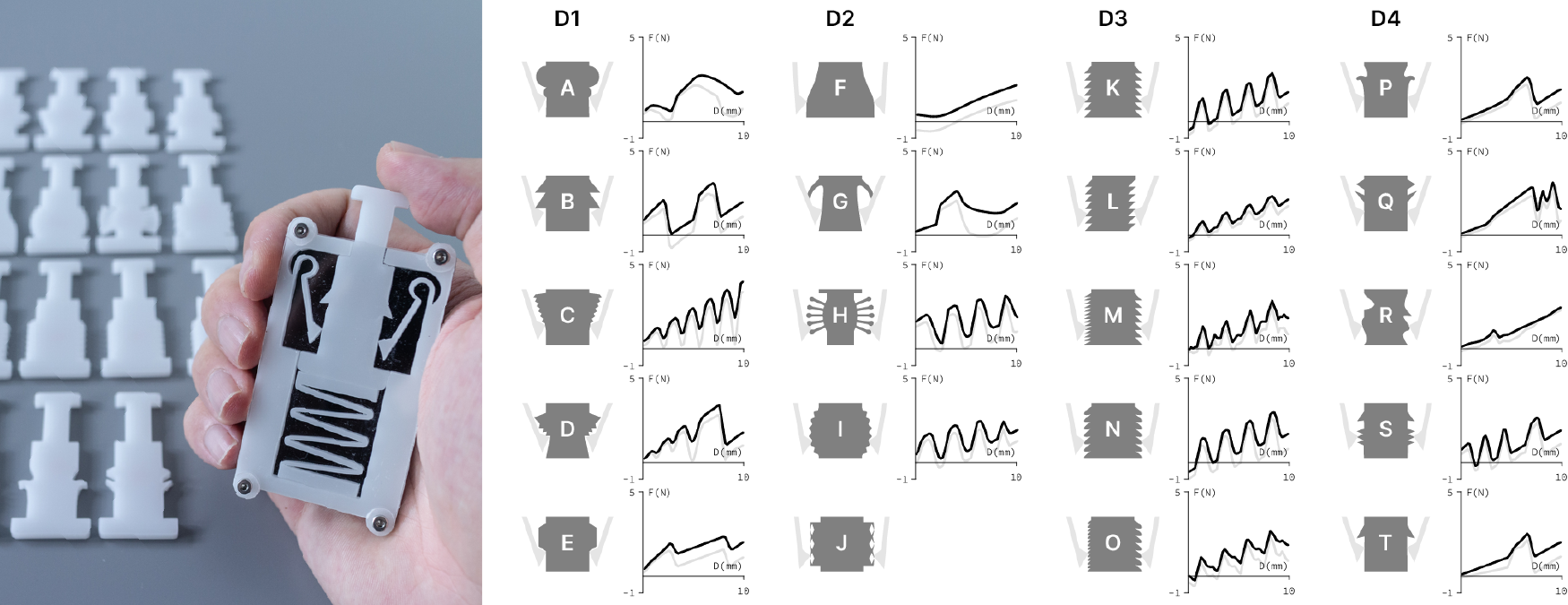}
    \Description[Pilot workshop outcomes]{20 outcomes of the pilot workshop from the 4 participants and their corresponding force-displacement curves.}
    \caption{Outcomes from the pilot workshop. Left: Fabricated outcomes. Right: Shortlisted mechanisms and their corresponding FD curves.}
    \label{fig:pilot}
\end{figure*}

\section{Pilot Design Workshop}

Besides our internal use, we piloted the sandbox with four professional industrial designers to assess its usability---and more importantly, how they might use it to explore and design haptic force feedback. All participants were involved in designing tangible interactive systems as part of their work (\Cref{tab:designers}).

\begin{table}[h]
\small
    \centering
    \begin{tabular}{L{0.55in} C{0.65in} L{1.7in}}
    \toprule
         
         & Professional Experience & Design Area \\ 
         
    \arrayrulecolor{black!30}\midrule
        
    D1 (female) & 2 years & Mixed reality medical simulators \\ 
    D2 (male) & 4 years & Physical medical simulators, medical devices \\ 
    D3 (male) & 5 years & Tangible interfaces, mixed reality interfaces \\ 
    D4 (male) & 4 years & Physical rehabilitation devices, tangible interfaces \\ 
    
    \bottomrule
    
    \end{tabular}
    \Description[Workshop participants]{Table of the workshop participants and their design area.}
    \caption{Workshop participants.}
    \label{tab:designers}
\end{table}

\subsection{Workshop structure}

The in-person workshop was conducted individually for each participant. In the first session, we provided physical demonstrations for participants to try and also walked them through the sandbox’s user interface and features. Participants were then tasked to design \name\ mechanisms via the sandbox and shortlist five designs to fabricate. To enable comparison across participants, we constrained their exploration to double-sided mechanisms as well as the same set of geometric parameters (\Cref{fig:ui}C). Participants were given a maximum of four days to complete the task (to respect their work schedule). We then fabricated their designs\footnote{We fabricated the workshop outcomes (rather than have participants fabricate them) in view of the safe management measures in place at our university due to the COVID-19 pandemic.} and conducted a semi-structured interview with them about their design process. In the interview, we asked them to recount their design process, as well as compare the fabricated samples to their expectations during digital design with the sandbox.

\subsection{Findings}

(We refer to specific task outcomes by their letter label in \Cref{fig:pilot}.)

\subsubsection{Outcomes} All participants completed the task and the outcomes are shown in \Cref{fig:pilot}. Out of the 20 designs, one design (J) could not be fabricated with the laser cutter as it had geometric features which were too thin. Checking for fabrication feasibility is a feature we intend to incorporate into future versions of the sandbox.

\subsubsection{Approaches} Participants tackled the open-ended task in a few different ways. D1 and D3 adopted a systematic approach; making incremental changes to the profile and observing the effect on the resultant FD curve. For instance, D3 started with a sawtooth pattern (K) which he called a ``control'' and made variations to that pattern; like doubling the frequency on one side (M) or rounding the teeth (N). D2 wanted to ``push the limits'' of the \name\ system and designed profiles that had compliant features as well (G, H). While the current FD curve estimation is not able to account for compliant-compliant interactions, this inspires further work. D4 contextualized the task and explored haptics for Nerf toy blasters---his personal hobby.

\subsubsection{Drawing with external tools} D2 and D3 designed their profiles in Adobe Illustrator and imported them into the sandbox. This feature was important to both of them and they mentioned that they preferred using a tool they were familiar with and had ``more control'' over. This flexibility enabled D2 to produce complicated profiles (e.g. H and J). 

\subsubsection{Visual-centric haptics design} D1, D3, and D4 made associations between the visual shape of the profile and the haptics they expect it will deliver. D4 described how he took a literal approach and designed the haptics for a trigger to visually resemble the shape of a trigger (P). D1 and D3 both discussed their intuition that curved profiles will have more ``gentle'' FD curves while sharp profiles will have ``sudden turns'' in the FD curve. \name\ mechanisms are visually obvious and support our participants to consider haptics from the appearance (shape) of the profile. However, it is important to highlight that ``what you see'' \textit{is not} ``what you get'' in terms of haptics---as evident in the estimated FD curves. Beyond the visual design of the profile, this pilot reveals the importance of facilitating haptic design via other modes which we discuss in the next few points.

\subsubsection{Range slider interaction} The range slider (\Cref{fig:slider}A) was highlighted by all participants as an essential interactive feature. D1 noted that the range slider ``let's me imagine how it might feel like when it is moved across a profile.'' D2 and D4 described how it supported their understanding of the FD visualization by comparing how the side spring interacts with a particular feature along the sliding profile with the corresponding point on the FD curve. Both D2 and D4 would adjust their profile designs to position haptic ``features'' along its displacement. For example, D4 adjusted the bumps along the profiles (P, Q, T) so that they encounter the side spring ``further down in the interaction [displacement]''.

\subsubsection{Reading FD curves} Participants made use of the FD visualization to different extents. D4 did not find the visualization helpful as he was unfamiliar with the concept. D1 and D3 mainly considered the amplitude of the FD curve. For instance, D1 will adjust a profile to maximize the peak force of the FD curve. The FD curves were helpful in challenging assumptions that participants had about haptic force feedback. D3 was surprised that the curved pattern (N) had a higher and more sustained peak value than the sharp pattern (K). D3 also made use of the compare feature (\Cref{fig:slider}B) to compare the FD curves between different designs, particularly against his ``control'' (K). Among the participants, D2 made the most use of the FD curves during his design process. For example, D2 had a goal of creating ``flat regions'' in the FD curve (i.e. regions of constant force) for profiles F and G. D2 described how he iteratively edited a profile while monitoring how the FD curve updates to achieve these constant force regions.

\subsubsection{Comparing physical mechanisms to FD curves} Interacting with the fabricated outcomes was the highlight of the workshop for our participants. Comparing the FD visualization to the physical outcomes, D1 noticed that she was ``less sensitive'' than she anticipated. In C for instance, she reported that the bumps did not feel like they were increasing in intensity. D3 noted that the FD curves for M versus K and O versus N did correspond to the haptic feedback provided by the fabricated objects---the asymmetrical profiles were ``dirty'', while the symmetrical profiles provided a ``cleaner'' haptic feedback. D3 also noted that the sharper profile K was more ``crisp'' compared to rounded profile N which felt more ``blunt''. D2 highlighted the importance of an iterative approach between digital design and physical fabrication for haptics design with the sandbox. D2 remarked that in a typical design process he would fabricate the object immediately after designing it and incrementally build his understanding of the design. For instance, after interacting with the fabricated outcomes, D2 noticed that fabricated mechanisms get stuck (i.e. stay in place midway instead of returning to their original position) at displacements where the visualized FD curve dips below zero. With that new knowledge, D2 mentioned that he will account for it the next time he uses the sandbox.

\subsubsection{Lowering the floor for haptic force feedback design} Through the interviews, all participants mentioned that haptics was often a secondary---or even arbitrary---consideration when they design physical interfaces. They mentioned that the sandbox was helpful in supporting inexperienced designers to explore passive force feedback mechanisms through \name. The observations from the pilot and the design processes that participants engaged in affirm the guidelines for haptic force feedback design tools proposed by \cite{van_oosterhout_facilitating_2020}. Specifically, the sandbox's import feature along with the built-in editor supported tool flexibility, while the range slider and FD visualization enabled visual previews to support a more systematic way to debug haptics. 

\subsubsection{Further studies and features} There are certainly more studies to be conducted to understand how we might further develop \name; for instance, we are interested to study how industrial designers use \name\ to design an entirely new object. This pilot also inspired us to think of how the sandbox can better support in navigating haptic force feedback via the FD curve. For instance, we can provide additional visual feedback on the FD curves that indicate just-noticeable-difference thresholds for human force perception \cite{feyzabadi_human_2013, prasad_force_2013, valverde_engineering_2019}. The sandbox can also be improved to enable bidirectional editing (i.e. authoring the FD curve and computing the resulting physical profile) to support designers to hone in on a target FD curve while exploring the physical shape of the physical profile.
\section{Applications}
\label{sec:propositions}

We built three types of interactive applications to showcase the breadth of the \name\ approach and demonstrate how industrial designers might use \name\ for designing physical interactive experiences.

\subsection{Haptic Attachments}

Digital fabrication is capable of rapidly prototyping bespoke physical parts on demand; and it can support designers to customize attachments that adapt the use existing objects and interfaces \cite{chen_reprise_2016}. \name\ mechanisms are planar and can be easily incorporated into existing objects to change their interaction haptics. We fabricated and applied such haptic attachment in three contexts:

\begin{figure*}[tbp]
    \centering
    \includegraphics[width=0.9\linewidth]{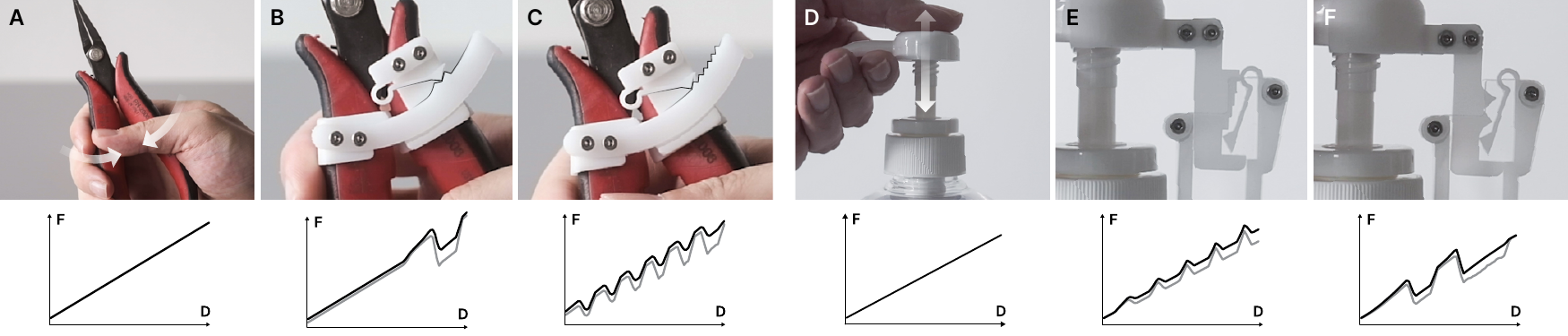}
    \vspace*{-2mm}
    \Description[Haptic attachments for pliers and pump bottle]{Shape-Haptics mechanisms were attached to a pair of pliers and a pump bottle. Plier A ``clicks'' when squeezed, while plier B has a racheting texture when squeezed. Pump bottle A has five distinct bumps when pressed, while pump bottle B has a strong mid point.}
    \caption{A, B, C: Pliers + haptic attachment and their corresponding FD curves. D, E, F: Pump bottle + haptic attachment and their corresponding FD curves.}
    \label{fig:objects}
    \vspace*{2mm}
\end{figure*}

\begin{figure*}[tbp]
    \centering
    \includegraphics[width=0.9\textwidth]{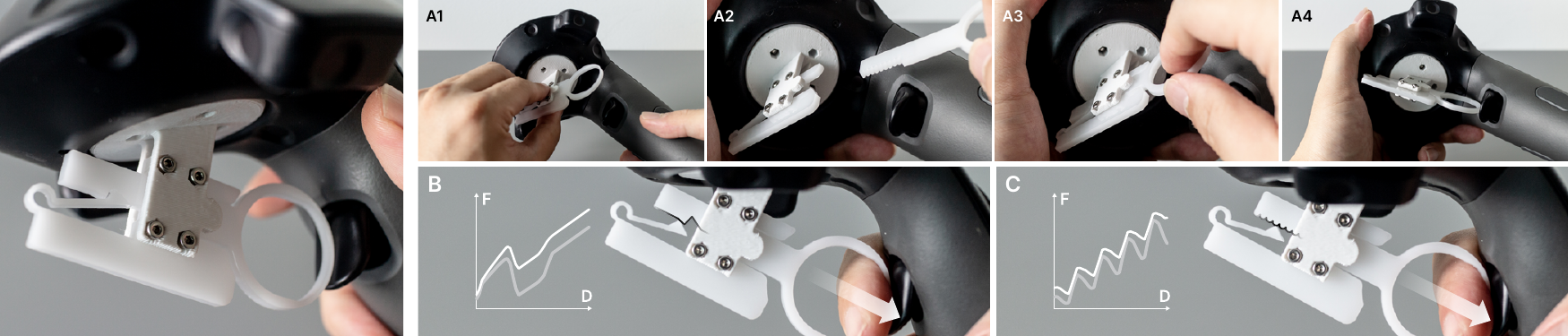}
    \vspace*{-2mm}
    \Description[Haptic attachments for VR controller]{Shape-Haptics mechanisms were attached to a Vive VR controller. The attachments are easily swapped in and out of the controller. Controller A has a mid point bump, while Controller B has a grating texture when pressed.}
    \caption{A1--4: Changing the sliding profile to modify haptics. B: Controller augmented with a midpoint ``click''. C: Controller augmented with a grating texture.}
    \label{fig:vr}
    \vspace*{2mm}
\end{figure*}

\begin{figure*}[tbp]
    \centering
    \includegraphics[width=0.9\textwidth]{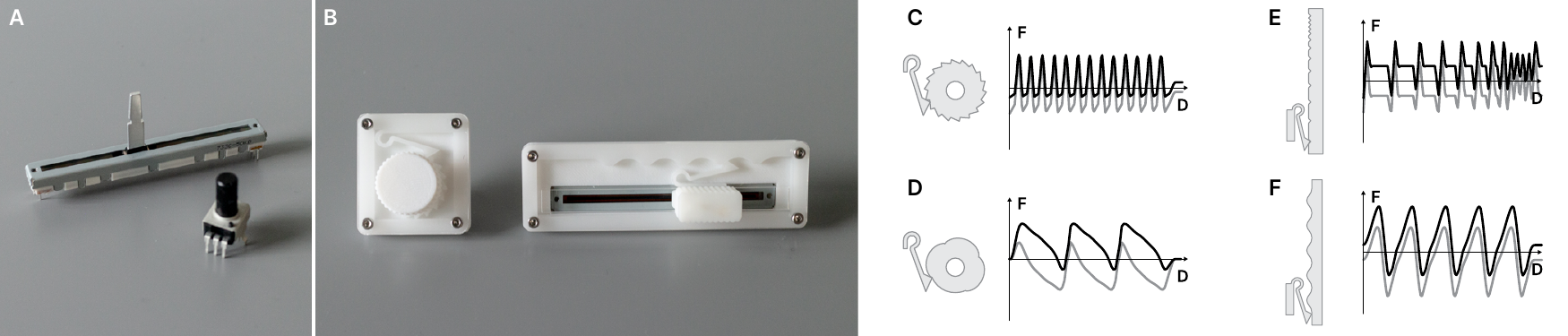}
    \vspace*{-2mm}
    \Description[Haptic attachments for electronic components]{Shape-Haptics mechanisms were attached to a knob and sliding potentiometer. Knob A has a ratcheting texture that feels different when rotated in different directions. Knob B has four stops separated by gentle bumps. Slider A has a series of detents that increase in frequency. Slider B has a wave profile along the track.}
    \caption{A: Off-the-shelf electronic knob and slider. B: Haptic attachments for knob and slider. C, D, E, F: Different sliding profiles and their corresponding FD curves.}
    \label{fig:electronics}
    \vspace*{2mm}
\end{figure*}

\subsubsection{For Everyday Objects}
We modified the haptic force feedback delivered by a pair of pliers and a pump bottle (\Cref{fig:objects}). The attachments leverage the existing spring in these objects. In the case of the pliers, we designed a mechanism that delivers a satisfying ``click'' right before it completely closes (\Cref{fig:objects}B), and another mechanism that delivers a directional haptic texture that is bumpy when squeezed and smoother on release (\Cref{fig:objects}C). In the case of the pump bottle, we explored using haptic attachments to provide a person with feedback about the amount they have pressed. The first mechanism delivers four distinct steps (\Cref{fig:objects}E) while the second mechanism delivers an obvious mid point (\Cref{fig:objects}F) when the pump is pressed. Besides augmenting the everyday objects in their existing context of use, such attachments can also be used as haptic proxies to augment interactions in a virtual environment. In addition they can be used as a brainstorming tool for product design; e.g. using the pliers + haptic attachment to explore haptic feedback for products with handles and levers like machine controls and surgical equipment.

\subsubsection{For VR controllers} 
Physical haptics improves the immersive experience and effectiveness of tools and objects in virtual reality \cite{insko_passive_2001, choi_claw_2018}. We developed a haptic attachment system for the Vive VR controller that enables designers to modify the haptics of the analog trigger. This system makes it easy to swap between different \name\ profiles (\Cref{fig:vr}A1--4); modifying the haptic force feedback of pulling the trigger. We fabricated two examples: a profile that delivers a ``click'' when the trigger is pressed halfway, and a profile with uniform bumps that delivers a grating texture.

\subsubsection{For electronic inputs}
Electronic input components such as push buttons, knobs (rotational potentiometers), and sliders (linear potentiometers) are typically used with a physical computing platform (e.g. Arduino, Raspberry Pi) to build physical interfaces. These input components conveniently instrument an interface with sensing functionality, but they also offer limited haptic feedback. We developed a variety of haptic attachments for an off-the-shelf knob and slider (\Cref{fig:electronics}B). By default, these components provide a constant frictional resistance when turned and pushed. For the knob, we fabricated a ratcheting profile that delivers different force feedback when rotated in different directions (\Cref{fig:electronics}C), as well as a notched profile that snaps at four locations along the knob’s rotation (\Cref{fig:electronics}D). For the slider, we fabricated a profile with indents arranged with increasing frequency (\Cref{fig:electronics}E), as well as a wave-like profile that ``pulses'' along its travel path (\Cref{fig:electronics}F).

\begin{figure*}[tbp]
    \centering
    \includegraphics[width=0.9\textwidth]{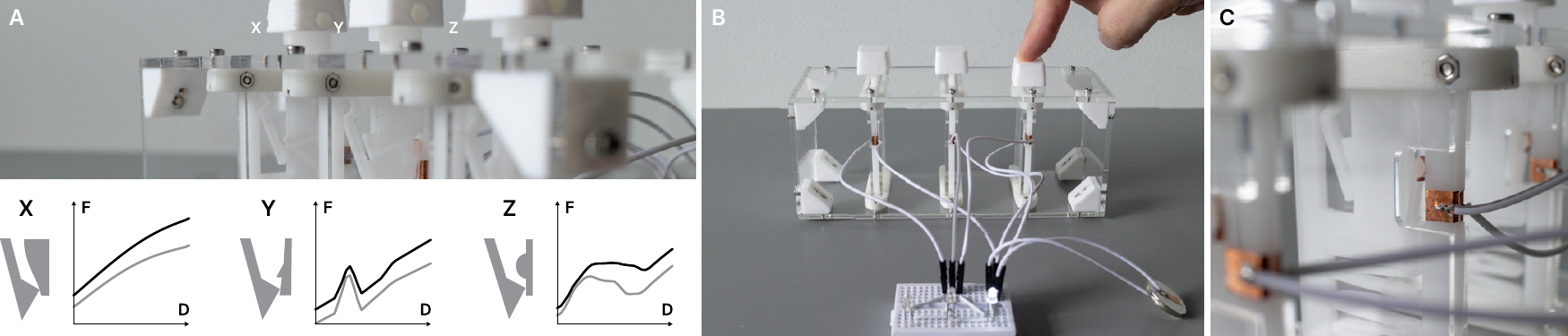}
    \vspace*{-2mm}
    \Description[Mechanical key tester]{Three different haptic mechanical keys were designed with \name. Key X offers a linear force feedback that tapers off. Key Y offers a sharp click at the beginning of the press. Key Z has a large bump in the middle of the press.}
    \caption{A: Three different haptic keys. B: Interacting with a key. C: Copper tape circuit with varying actuation points.}
    \label{fig:keytest}
    \vspace*{2mm}
\end{figure*}

\begin{figure*}[tbp]
    \centering
    \includegraphics[width=0.9\textwidth]{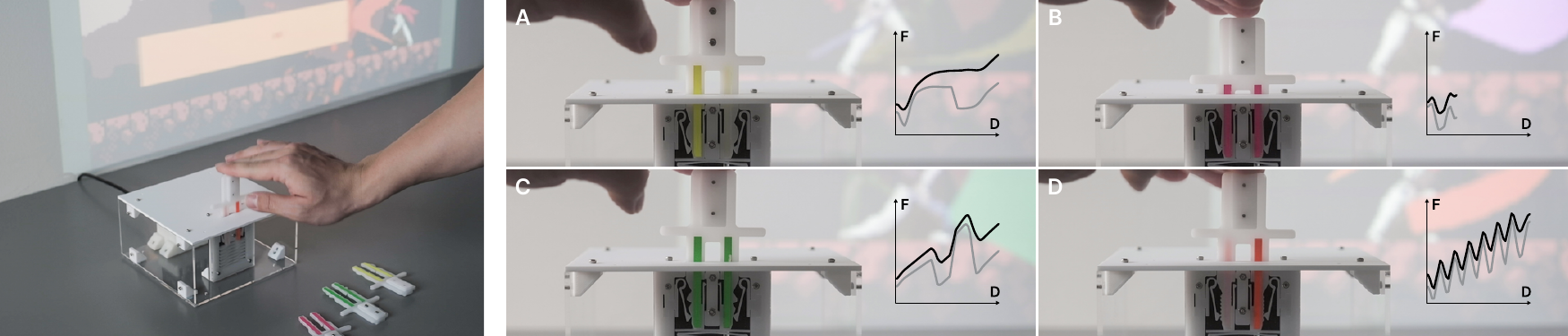}
    \vspace*{-2mm}
    \Description[Switchblade video game controller]{Switchblade offers four physical inputs with different Shape-Haptics profiles that can be swapped in and out of a custom video game controller. Each physical input therefore offers different haptic feedback.}
    \caption{\textit{Switchblade} game controller with swappable inputs.}
    \label{fig:switchblade}
    \vspace*{2mm}
\end{figure*}

\subsection{Mechanical Key Tester}

In this example, we devised a setup for mechanical keyboard enthusiasts to explore different haptic profiles for computer keys. We constructed a rig with three linear \name\ mechanisms presented as keys (\Cref{fig:keytest}A). Sliders with different profiles can be swapped into this rig to deliver different force-feedback profiles. We also modified the basic \name\ mechanism to include a copper tape contact switch which closes when the button is pressed. We were inspired by the button experiments conducted in \cite{liao_button_2020} and designed the switch so that its actuation point can be easily adjusted by shifting the location of its copper tape contacts. This could be used to evaluate the efficacy of a button for performing a task that is dependent on reaction time, such as in video gaming.

\subsection{Alternative Game Controller}
We developed \textit{Switchblade}---a video game with a bespoke physical controller designed with \name. Switchblade is a rhythm game where the player controls a swordsperson. The game controller comprises one socket and the player has to switch between four different blades to match the color of the virtual enemies on screen. The physical blades have different sliding profiles as well as travel displacements; and pressing each of them provides different haptic feedback. We designed the force feedback of each blade to match the semantics of the in-game action---for example a quick draw sword delivers a ``swooping'' haptic profile (\Cref{fig:switchblade}A), a heavy sword requires a harder two-step push (\Cref{fig:switchblade}C), and a set of twin daggers that slice rapidly delivers ``jagged'' force feedback (\Cref{fig:switchblade}D). The swappable inputs are read by the controller via a simple computer vision script that detects the blade’s color. Through Switchblade, we demonstrate a novel input system with interchangeable haptic force feedback with simple passive parts.
\section{Limitations \& Opportunities}
\label{sec:limitations}

\subsection{Limitations}

\subsubsection{Laser cutting} 
Laser cutting is a rapid prototyping process capable of fabricating intricate 2D patterns; and it is central to \name. However, laser cutters produce parts with a margin of dimensional error due to potential inconsistencies in the laser cut kerf and surface irregularities along the laser cut edge. These introduce inconsistent friction into the mechanisms. To manage these fabrication limitations, we ensure that a part's geometric details are well within the laser cutter's tolerance (\Cref{fig:laser}). We also visually inspect parts to check for obvious blemishes along the laser cut edges which are removed with manual filing.

\subsubsection{Force feedback range}
For this paper, we focused on mechanisms that deliver light force feedback in the range of 0--5N. This is typically associated with haptics for hands/fingers. This is largely due to the specific material (3mm thick POM sheets) and fabrication process (laser cutting) we adopted. Future work includes extending the concepts presented in this paper to other materials and fabrication processes, such as thicker POM sheets or metal, to accommodate heavier force feedback.

\subsubsection{Movement types}
We focused on one-dimensional linear and rotary mechanisms in this paper; a subset of movement types for physical interfaces \cite{mackinlay_semantic_1990}. While we demonstrate that these simple movements can be extended and applied to a variety of applications (such as using a linear mechanism to activate the angular trigger movement of a VR controller, \Cref{fig:vr}), there is an opportunity extend the simple mechanisms described here and explore more complex passive force feedback mechanisms that move differently or with multiple degrees of freedom.

\subsubsection{Haptic force feedback estimation}
The estimated FD curve is a first---albeit rudimentary---step to explain the haptics delivered by passive force feedback systems. It does not capture other factors that contribute to haptics, such as vibrations and interaction velocity \cite{liao_button_2020}. The FD estimates provided by the sandbox aim to provide an initial layer of information to facilitate designing mechanisms that will ultimately be fabricated and experienced physically. With that in mind, there is an opportunity to incorporate the concepts presented by \cite{liao_button_2020} to provide more nuanced FD estimations during the digital design process.

\subsection{Opportunities}

\subsubsection{Passive physical tuning}
When assembling the plier + haptic attachment (\Cref{fig:objects}), we noticed that loosening the screws on the side spring introduces ``play'' during interaction---vibrations in the part alongside the spring's compression. These movements happen along with the compression and release of the side spring and results in more nuanced haptic feedback. Anecdotally, the research team felt that the system with more play had a sharper ``click'' compared to the tighter system. This inspires us to look into tuning as simple means of modifying the passive force feedback of a fabricated mechanism (akin to tuning the strings of a musical instrument).

\subsubsection{Compliant on compliant systems} 
\name\ mechanisms comprise a compliant side spring that engages with a rigid slider to deliver force feedback. During our pilot however, D2 raised the possibility of incorporating compliant structures into the sliding profile for more complex haptic force feedback (\Cref{fig:pilot}H). We believe this will further expand the range of haptic expressions with such mechanisms---and plan to explore them in future work.

\section{Conclusion}
This paper traces our research into designing haptics for physical interfaces through passive force feedback mechanisms. We took multiple lines of inquiry in this work, including exploring materials and digital fabrication, computational design, designer studies, and building interactive demonstrations. Through this process, we provided data and insight into laser cutting compliant spring structures with POM. We also proposed a simple and efficient way to estimate the force-displacement curves of \name\ mechanisms and developed a computational design sandbox to facilitate exploring and fabricating these mechanisms. 
Through the variety of haptic mechanisms we created and tested in our material exploration, as well as from the workshop outcomes, we see potential in \name\ as a design approach for supporting a wide range of passive haptic expressions.
Professional designers who participated in the workshops were able to jump into the \name\ sandbox quickly and explore haptic force feedback with it. These initial findings echo \cite{seifi_how_2020, van_oosterhout_facilitating_2020, seifi_haptipedia_2019} and highlight the importance of such platforms for lowering the floors to enable more people to engage with haptic design.
Furthermore, with the application examples we showcase how designers can extend \name\ to build haptic physical interactions that cater to different design scenarios.

We hope that these contributions, in part, will inspire HCI researchers working in the areas of materials, fabrication, and designing tangible interactions. And as a whole, we hope to offer a new perspective on how passive haptics can be explored, designed, and built for both new---and existing---physical interfaces.

\begin{acks}
We would like to thank Han Bo, Vivien Tan, Luke Goh, and Mok Zijie from the Division of Industrial Design at National University of Singapore for their assistance during research. We would also like to thank our workshop participants and the review team for their suggestions and feedback.
\end{acks}

\balance
\bibliographystyle{ACM-Reference-Format}
\bibliography{bib}

\end{document}